\documentclass[rnote]{aa}
\usepackage{graphicx}
\usepackage[english]{babel}
\usepackage[varg]{txfonts}
\usepackage{multirow}
\usepackage{lscape}
\usepackage{longtable}
\usepackage{natbib}
\usepackage{color}
\usepackage{float}
\usepackage{url}
\newcommand{\OII}{{[O\textsc{ii}] }}
\newcommand{\OIII}{{[O\textsc{iii}] }}

\begin{document} 

   \title{Measuring galaxy \OII emission line doublet with future ground-based wide-field spectroscopic surveys}
   \titlerunning{Resolution and \OII redshift surveys}

    \author{Johan Comparat,\inst{1} \and 
Jean-Paul Kneib\inst{2,1} \and  Roland Bacon\inst{3} \and Nick J. Mostek\inst{4} \and Jeffrey A. Newman\inst{5} \and David J. Schlegel\inst{4} \and Christophe Y\`eche\inst{6} }
\institute{
Aix Marseille Universit\'e, CNRS, LAM (Laboratoire d'Astrophysique de Marseille) UMR 7326, 13388, Marseille, France\\ \and
Laboratoire d'astrophysique, \'Ecole Polytechnique F\'ed\'erale de Lausanne (EPFL), Observatoire de Sauverny, 1290 Versoix, Switzerland\\ \and
 CRAL, Observatoire de Lyon, Universit\'e Lyon 1, 9 Avenue Ch. Andr\'e, 69561 Saint Genis Laval Cedex, France\\ \and
 Lawrence Berkeley National Laboratory, 1 Cyclotron Road, Berkeley, CA 94720, USA\\ \and
 Department of Physics and Astronomy, University of Pittsburgh and PITT-PACC, 3941 O'Hara St., Pittsburgh, PA 15260, USA \\ \and
 CEA, Centre de Saclay, IRFU, F-91191 Gif-sur-Yvette, France\\
}

\date{Accepted by A\&A on sept. 20th 2013}

\abstract{
The next generation of wide-field spectroscopic redshift surveys will map the large-scale galaxy distribution in the redshift range $0.7\leq z\leq2$ to measure baryonic acoustic oscillations (BAO). The primary optical signature used in this redshift range comes from the \OII emission line doublet, which provides a unique redshift identification that can minimize confusion with other single emission lines.
To derive the required spectrograph resolution for these redshift surveys, we simulate observations of the \OII ($\lambda\lambda$ 3727,3729) doublet for various instrument resolutions, and line velocities. 
We foresee two strategies about the choice of the resolution for future spectrographs for BAO surveys. 
For bright \OII emitter surveys (\OII flux $\sim 30\times 10^{-17} \mathrm{erg \, cm^{-2} \, s^{-1}}$ like SDSS-IV/eBOSS), a resolution of $R\sim3\,300$ allows the separation of 90 percent of the doublets. The impact of the sky lines on the completeness in redshift is less than 6 percent. 
For faint \OII emitter surveys (\OII flux $\sim 10\times 10^{-17} \mathrm{erg \, cm^{-2} \, s^{-1}}$ like DESi), the detection improves continuously with resolution, so we recommend the highest possible resolution, the limit being given by the number of pixels (4k by 4k) on the detector and the number of spectroscopic channels (2 or 3).}

\maketitle 
\section{Introduction}
\label{section:introduction}
Following the successful baryonic acoustic oscillation (BAO) measurement in the galaxy clustering from SDSS \citep{Eisenstein_2005}, 2dFGRS \citep{2005MNRAS.362..505C}, Wiggle-Z \citep{blake2011}, and the Baryonic Oscillation Spectroscopic Survey (BOSS) \citep{2012MNRAS.427.3435A}, there is a strong motivation in the community to plan the next generation of spectroscopic redshift surveys for BAO. In particular, the future ground-based surveys plan to map the galaxy distribution in the redshift range $0.7\leq z\leq2$ and use the galaxy power spectrum to precisely measure the BAO signature and constrain the cosmological parameters.  

Two examples of this new paradigm are the following projects: SDSS-IV/eBOSS and DESi.
The SDSS-IV/eBOSS dark energy experiment starts observing in 2014 with SDSS-III/BOSS infrastructure ($1\,000$ fibers on $\sim$7 deg$^2$). This survey will measure about 1.5 million spectroscopic redshifts of QSOs in the redshift range $0.9<z<2.5$ and galaxies with a redshift in $0.6<z<1.2$.
The DESi project plans to map 14$\,$000 deg$^2$ of sky using 5$\,$000 motorized fibers over a 7 deg$^2$ field of view and to measure 22 million galaxy redshifts; see \citet{bigBOSS_2011} for a global survey description and \citet{2012SPIE.8446E..0QM} for the current survey parameters. 

Galaxy redshifts will be mostly determined from the emission line features of star-forming galaxies between $0.7\leq z\leq2$. Table \ref{talbe:oii} lists the primary emission lines that are available at optical and NIR wavelengths within this redshift range. Of these lines, the \OII  doublet at $(\lambda 3727,\lambda 3729)$ will provide the most consistently available feature. In order to avoid confusion with other prominent emission lines (H$_\alpha$, H$_\beta$, \OIII), the \OII doublet should be resolved over the instrumented wavelength range where no other lines are available to make an unambiguous identification.

\begin{table}
	\caption{Emission lines available at optical and near-infrared wavelengths. Taken from Atomic Line List (from www.pa.uky.edu). $\lambda_\mathrm{vac}$ is the wavelength emitted in vacuum in $\AA$, the orbital transition is given under the column `term', `J-J' is the spin state. The last column gives the energy transition that occurs in electron-Volt.}
	\label{talbe:oii}
	\centering
\begin{tabular}{c c c c c}
\hline \hline
line name & $\lambda_\mathrm{vac}$  & J-J & energy levels  \\
 		& $(\AA)$						&	&	(eV) \\
\hline 
\OII & 3$\,$727.092    &    3/2 - 3/2 & 0 - 3.326 \\
\OII & 3$\,$729.875    &    3/2 - 5/2 & 0 - 3.324 \\
$H_\beta$ & 4$\,$862.683 &  * - * & 10.198 -   12.748\\
\OIII & 4$\,$960.295    &        1-2      &  0.014 -    2.513 \\
\OIII & 5$\,$008.240   &      2 - 2   & 0.037 -    2.513 \\
$H_\alpha$ & 6$\,$564.61 &  * - * & 10.198 -   12.087\\

\hline
\end{tabular}
\end{table}

Previous emission line redshift surveys have had different strategies concerning the use of emission lines for measuring the redshift. Wiggle-Z with a spectral resolution of 1300 obtained 60\% of reliable redshifts (18\% based on the detection of the \OII doublet {\it i.e.} the doublet is resolved or partially resolved), and 40\% of unreliable redshifts; \citep{2010MNRAS.401.1429D}. DEEP2 survey with a resolution of 6$\,$000 obtained 71\% of reliable redshifts (14.8\% based on the detection of the \OII doublet {\it i.e.} the doublet is resolved or partially resolved), 10\% between reliable and unreliable, and 19\% of unreliable redshifts \citep{2012arXiv1203.3192N}. 
The difference between these redshift efficiencies is related to the resolution of the spectrograph and the wavelength it covers.
Indeed, if the \OII emission is the only one available in the spectrum, at high resolution the doublet is split and the redshift is reliable. Whereas, at lower resolution the \OII doublet is not always split and may be taken for another emission line.

In Section \ref{sec:req}, we derive the minimum resolution necessary to resolve the doublet in the case of an observation without noise. In Section \ref{sec:simulation}, we describe our simulation of \OII doublet detections based on DEEP2 spectral observations. We discuss the results of our simulation in Section \ref{sec:result}.
\section{Instrumental requirements}
\label{sec:req}
First, let us define our notation. $R=\lambda/FWHM_\lambda$ is the resolution of the spectrograph, $\lambda_a=3\,727.092\AA$, $\lambda_b=3\,729.875\AA$ are the individual \OII emission wavelengths and $\lambda_{\OII}=(\lambda_a*3.326568+\lambda_b*3.324086)/(3.326568+3.324086)=3\,728.483\AA$ is the energy-weighted mean \OII wavelength. The observed wavelength separation between the emission lines depends on the redshift $\delta_{[O\textsc{ii}]}(z)=(\lambda_b-\lambda_a)(1+z)=2.783(1+z).$

We can thus define the resolution, $R_{\OII}$, as the minimal resolution required to properly sample a theoretical \OII doublet (with zero intrinsic width) without loss of information by $R_{[O\textsc{ii}]}=2{(1+z) \lambda_{[O\textsc{ii}]}/\delta_{[O\textsc{ii}]}(z)}= 2\,679$ (Nyquist-Shannon sampling theorem, \citealt{1975mtcbookS}). Note that $R_{\OII}$ is independent of redshift.

However, a {\sl real} galaxy has an intrinsic velocity dispersion, $\Delta v$, that broadens the emission lines from 
a theoretical Dirac $\delta$-function profile. Assuming the line profile is dominated by thermal Doppler broadening
in the host galaxy interstellar medium,
the observed wavelength width, $\delta \lambda_v$, of the broadened \OII line profiles is defined in Eq. \ref{eq:doppler} where $c$ is the speed of light.
\begin{equation}
\delta \lambda_v =  \lambda_{\OII} \frac{\Delta v}{c}.
\label{eq:doppler}
\end{equation}
In this simplified case, the intrinsic velocity dispersion is equivalent
to the standard deviation in a Gaussian profile. For example, a galaxy at $z=1$ with $\Delta v=50 \, \mathrm{km\, s^{-1}}$ has a line width of $\delta \lambda_v\sim0.6 \AA$, which represents $\sim10\%$ of the wavelength separation between the doublet peaks.

Furthermore, the spectral resolution of the instrument also broadens the width of the \OII lines. The change in line width due to resolution is given by $\delta \lambda_R(z)$ defined in Eq. \ref{eq:resolution}. Note that the broadening due to instrumental resolution depends on the redshift because the position of \OII changes with redshift while the resolution element $FWHM_\lambda$ is roughly constant with the wavelength (for a grism spectrograph):
\begin{equation}
\delta \lambda_R(z) =  (1+z) \frac{\lambda_{\OII}}{R}.
\label{eq:resolution}
\end{equation}
By performing a squaring sum of the components in Eq.~\ref{eq:doppler} and \ref{eq:resolution}, we obtain the observed width, denoted $w_{\OII}(z)$, of an individual line in the \OII doublet:
\begin{equation}
w_{\OII}(z) =  \lambda_{\OII} \sqrt{  \frac{(1+z)^{2}}{R^{2}}+ 
					 \frac{\Delta v^{2}}{c^{2}} } .
\label{eq:doublet}
\end{equation}
In order to Nyquist sample the observed \OII doublet at redshift $z$, the individual line width of the doublet has to be at least twice the doublet separation, or $w_{[OII]}(z) = 2 \delta_{\OII}(z)$.
Rewriting Eq. \ref{eq:doublet} in terms of this minimum sampling requirement gives the minimal resolution, denoted $R(z,\Delta v)$, required to split an \OII doublet emitted at redshift $z$ with a velocity dispersion $\Delta v$:
 \begin{equation}
R(z,\Delta v)  = \left[  \frac{1}{R_{\OII}^{2}}- 
					 \frac{\Delta v^{2}}{(1+z)^{2}c^{2}} \right]^{-1/2}.  
\label{eq:6:Rzv}
\end{equation}
$R(z,\Delta v)$ decreases with redshift, increases with the velocity dispersion, and it converges asymptotically towards $R_{\OII}$.

For a galaxy at $z=1$ (\OII is observed at $\lambda\sim7456\AA$) with $\Delta v=100\, \mathrm{km\, s^{-1}}$, the minimum resolution required is $R_{min}=3\,000$. For a galaxy at $z=1$ with $\Delta v=70\, \mathrm{km\, s^{-1}}$, it is $R_{min}=2\,800$. The spectrograph currently used by SDSS-III/BOSS reaches $R\sim2700>R_{\OII}$ at $9\,320\AA$, and therefore it theoretically splits the \OII doublet for galaxies with $\Delta v<50\, \mathrm{km\, s^{-1}}$ at $z\geq1.5$. With this spectrograph, the observation of the \OII doublet of galaxies with $\Delta v=100\, \mathrm{km\, s^{-1}}$ will be highly-blended. 

At low resolution, it is possible to actually see \OII doublets when the lines peaks and valley falls exactly right opposite to the pixels. In the following, when we state `the doublet is resolved', it is true wherever the emission line lands on the detector.

In classical spectrographs, the resolution element $FWHM_\lambda$ is roughly constant with wavelength, and therefore the spectral resolution $R$ is a linear function of  the observed wavelength $\lambda$.  
We must therefore define the minimum resolution requirement to be at the lowest redshift limit where  \OII becomes the only emission line available in the spectrum. The resolution requirement will automatically be satisfied for all higher redshifts.


In this study, we consider the more common case where the spectral range is limited to $<1\mu m$. 

\section{Simulation}
\label{sec:simulation}

To confirm the theoretical considerations of Section 2, we simulate observations of \OII doublets in the presence of Poisson noise. Future massive spectroscopic redshift surveys are primarily focused on obtaining redshifts with only emission lines, which is less demanding in terms of exposure time than requiring the detection of the continuum. For these applications, a Gaussian profile is sufficient to simulate the resolution effects.

Because spectrograph resolution increases with wavelength, the minimal resolution requirement is determined at the shortest wavelength where the \OII doublet becomes the only major emission line feature in the spectrum. Assuming an instrumental wavelength limit of 1$\mu$m, the resolution requirement for \OII is therefore defined at $\lambda_{obs}(\OII,z=1)\sim 7\,450$\AA.

Of interest for this work, DEEP2 has obtained a complete spectroscopic 
sample of \OII emitters at redshift $z=1$. Its magnitude limit is $r=24.1$ and its \OII flux limit is $5\times 10^{-17} \mathrm{erg \;cm^{-2}\;s^{-1}}$ \citep{2012arXiv1203.3192N}. These limits are deeper than the target selection limits for BAO surveys currently under development. DEEP2 used the DEIMOS grism spectrograph at Keck with a resolution R=6$\,$000 \citep{2003SPIE.4841.1657F} and
was limited by the galaxy continuum signal-to-noise. 

The range of velocity dispersions used in our simulation is empirically determined by observations of $z\sim1$ \OII emitters within the DEEP2 redshift survey. We set the lower (upper) limit of the investigated range at $\Delta v=20 \, \mathrm{km\, s^{-1}}$ ($120 \, \mathrm{km\, s^{-1}}$), which encompasses most of the galaxies down to $r<24$.

In terms of instrumental resolution, we explore the range of  $2\,500<R<6\,000$ sampled by steps of $\delta_{R}=3$ in resolution. To avoid aliasing problems, for each doublets we add a random number smaller than 3 to the resolution, in order to sample correctly the complete resolution range. 
We use a sampling of 3 pixels per resolution element. 
Our results will span a meaningful range of resolutions for numerous spectrographs at $\lambda_{obs}(\OII)\sim 7\,500$\AA, including the current SDSS-III/BOSS spectrograph ($R\sim2\,500$, \citet{2013AJ....146...32S}) and future spectrographs such as PFS-SUMIRE ($R\sim3\,000$ \citet{2012SPIE.8446E..4PV}) or DESi ($R\sim4\,000$ \citet{2012SPIE.8446E..68J}).

We use a Gaussian function, to model the \OII doublet, given by 
$f_\mathrm{gaussian}(\lambda,\lambda_0,\sigma_g,F_0)= \frac{F_0}{\sqrt{2\pi} \sigma_g}  Exp \left[ \frac{(\lambda-\lambda_0)^2}{2 \sigma_g^2} \right]$.
This produces an emission line centered at $\lambda_0$ of total flux $F_0$. The profile width $\sigma_g$ is linked to the velocity dispersion by $\sigma_g=\lambda_{\OII} \Delta v / c$. The Gaussian profile has an exponential drop off from the emission peak value, and therefore it may not represent systematic effects like scattered light within the spectrograph. A Moffat profile recovers the information in the wings of the emission line when $\beta$ is allowed to vary. However, the Moffat model is only attractive if the data has a high spectral resolution and high signal to noise ratio. Otherwise, the information in the wings will have low significance due to measurement noise.

We calibrate the flux $f$ and the sky level to a recent emission line galaxy observational study performed at the SDSS Telescope \citep{2013MNRAS.428.1498C}. This study showed the nominal observed total line flux is $\sim 30\times 10^{-17} \mathrm{erg \, cm^{-2} \, s^{-1}}$ and the nominal sky brightness is $\sim 3\times 10^{-18} \mathrm{erg \, cm^{-2} \, s^{-1} \, \AA^{-1} arcsec^{-2}}$ at $\sim 7400 \AA$. This noise level corresponds to detections with a SNR above 7 which should be typical of observations in future BAO survey. In the simulation, we use fluxes $f$ from a broader range, $6<f<100\times 10^{-17} \mathrm{erg \, cm^{-2} \, s^{-1}}$. We determine the relative abundance of emission lines at a given flux with the \OII luminosity function at $z\sim1$ measured by \citet{2009ApJ...701...86Z} on DEEP2 survey.

First, we first make a Gaussian doublet at $\lambda_{obs}(\OII)\sim 7450\AA$ for a given resolution $R$, velocity dispersion $\Delta v$ , and flux $f$. The flux ratio between the two lines is fixed at 1, the impact of a varying flux ratio is discussed in the paragraph \ref{subsec:fluxratio}. 
Next, we sample the doublet spectrum at resolution $R$ with 3 pixel per resolution element. We add Poisson sky noise on each pixel (this is the dominating contribution of the observed noise). This creates a mock observation of the \OII emission doublet for the Gaussian profile.
Finally, we fit two models to the simulated doublet: a single Gaussian profile, and a double Gaussian profile. From each fit, we compute the $SNR$ and the $\chi^2$ to compare the detections. $\chi^2$ is defined as the usual `reduced chi-square statistics' by 
$\chi_{i=1\,or\,2}=\frac{1}{n_{dof}} \sum_{k \; \in\; \mathrm{pixels}} \frac{(O_k-M^{i}_k)^2}{N_k^2}$
where $n_{dof}$ is the number of degrees of freedom, $O$ is the array of observed values, $M^1$ is the model with one line, $M^2$ is the model with 2 lines, and $N$ is the noise. The number of degrees of freedom vary from 35 to 94 (depending on the spectral resolution used). 
The $SNR$ is calculated with a Fisher matrix.

\section{Results}
\label{sec:result}
The simulation contains $\sim15\times10^6$ simulated \OII lines sampling the velocity dispersion, resolution, and flux range set in the above. 

\begin{figure}
\begin{center}
\includegraphics[width=88mm]{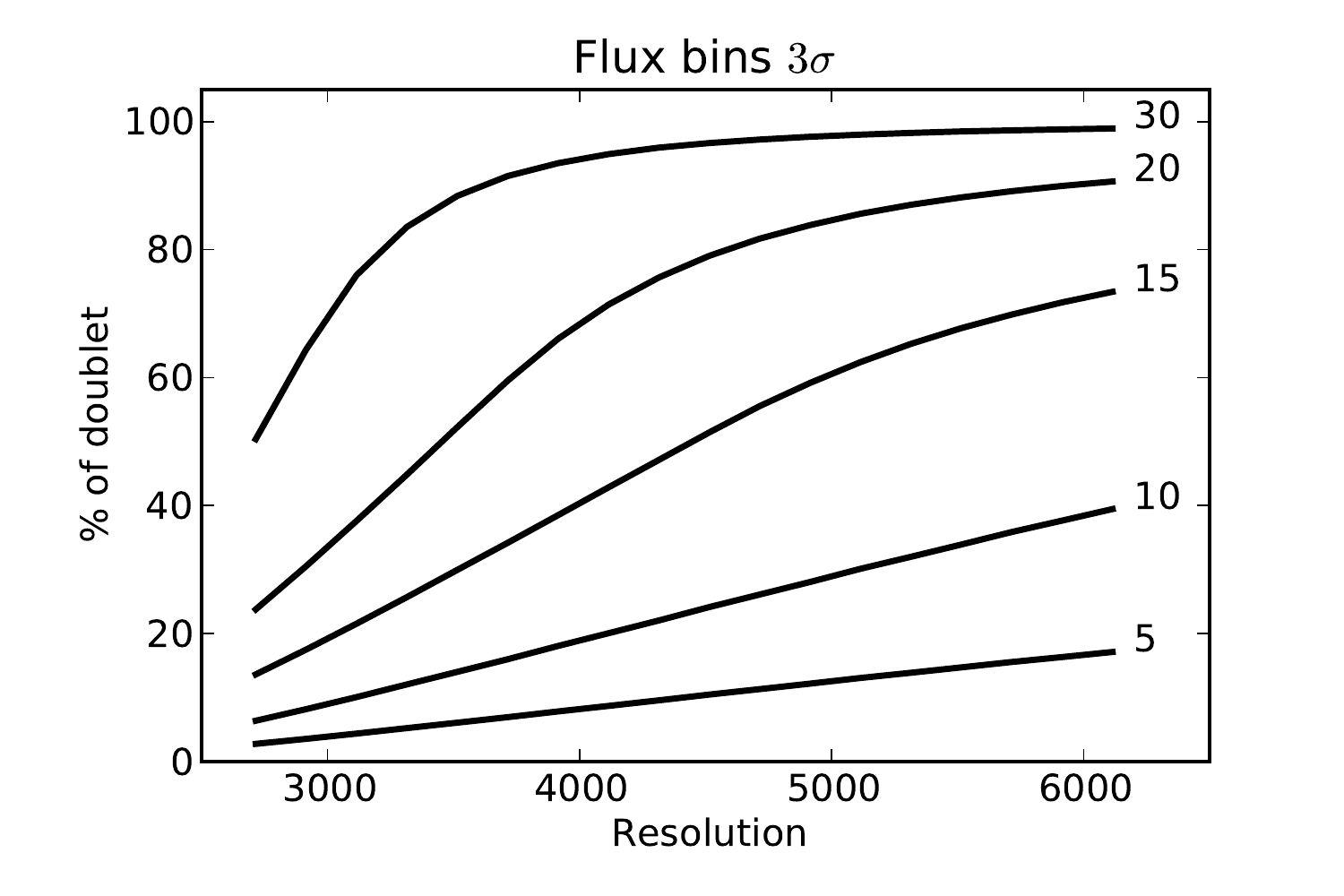}
\caption{Share of doublets at the 3$\sigma$ (confidence level of 99.7\%) vs. resolution for $r<24$ doublets at $z=1$ for different flux bins and with a flux ratio between the lines of 1. Each line corresponds to a survey with a the flux detection limit given on the right end of each line in units of $10^{-17}\mathrm{erg \, cm^{-2} \, s^{-1}}$. SDSS-IV/eBOSS corresponds to the line 30 and DESi to the line 10.} \label{figure:letter4}
\end{center}
\end{figure}

To statistically differentiate whether an observation of \OII is identified as a doublet or a single emission line (SEL), given that the numbers of degrees of freedom is high ($35<n_{dof}<94$), we use the difference $\Delta \chi^2=\chi_1/n_{dof1} - \chi_2/n_{dof2}$ of the normalized $\chi^2$. A $\Delta \chi^2=9$ means the single line emission model is ruled out at $3\sigma$ or with a 99.7\% confidence level. We compute the share of emission line with $r<24$ (convolved by the velocity dispersion distribution of DEEP2) detected as a doublet at the 3 $\sigma$ confidence levels at redshift 1 as function of the resolution for different \OII flux detection limit, see Fig. \ref{figure:letter4}.

The main trend is that the percentage of doublets increases as a function of the resolution. We can distinguish two regimes.
In the regime of low \OII fluxes the gain is linear, {\it i.e.} for surveys with a lower limit of \OII detection of 10$\times10^{-17}\mathrm{erg \, cm^{-2} \, s^{-1}}$ or below the increase of the share of doublet is linear as a function of the resolution (it corresponds to the line 10 of Fig. \ref{figure:letter4}). For such survey, it indicates the resolution should be the highest possible.
For higher \OII fluxes, the marginal increase of the doublet share is large for low resolutions and small for higher resolution.
For a survey aiming only to observe the brightest \OII emitters (on Fig. \ref{figure:letter4}), it is not necessary to aim for the highest resolution. $R=3\,300$ is sufficient to obtain 90\% of doublets. And for $R>3\,300$, the marginal cost of an extra percent of doublets decreases.

The DEEP 2 survey dealt with SEL using a neural network \citep{2007ApJ...660...62K}. They showed that given a fair spectroscopic sample of an observed population with reliable redshifts, it is possible to infer correct redshifts to nearly 100\% of the \OII SEL. The $H_\alpha$, $H_\beta$, and \OIII SEL cases are not as well handled by the neural network with efficiencies of $\sim90\%$, $\sim60\%$, and $\sim60\%$ respectively.

The combination of the two latter points shows it will be possible to derive robust \OII redshifts where \OII is the only emission line available in the spectrograph, even if the fraction of $3\sigma$ doublet detections is small.

\subsection{Higher redshift, sky lines, completeness}
The sky lines have an observed width of one resolution element, therefore their width varies with the resolution. In the case of a single sky line located on a doublet, it is not a problem to subtract the sky line and obtain an accurate redshift. In the case of many contiguous sky lines, it can cover completely a doublet and prevent from getting any redshift in this zone. This causes the survey to have a varying \OII flux limit as a function of the redshift. To quantify the impact of the sky lines obstruction as a function of redshift, we simulate at various resolutions the observation of a sky spectrum. The sky spectrum is taken from \citet{2003A&A...407.1157H}. 

At a given resolution, we convert the wavelength array of the sky into a redshift array corresponding to the \OII redshift. We scan the redshift array by steps of 0.0005 (it corresponds to the desired precision of a spectroscopic redshift). At each step, we compare the median value of the sky (we assume here a sky subtraction efficient at 90\%) to the flux measured in the middle of an \OII doublet with (where it is the lowest). If the median value of the sky is greater than the value of the doublet, we consider we cannot fit a redshift. Finally, we compute the percentage of the redshift range where we can fit spectroscopic redshifts. 

We run this test for two settings. A bright survey with \OII fluxes $\sim 30\times 10^{-17} \mathrm{erg \, cm^{-2} \, s^{-1}}$ and fibers of 2'' diameter (SDSS-IV/eBOSS-like). A faint survey with \OII fluxes $\sim 10\times 10^{-17} \mathrm{erg \, cm^{-2} \, s^{-1}}$ and fibers of 1.5'' diameter (DESi-like). 
It shows the impact of the sky lines on the completeness depends weakly on the resolution, the discrepancy between the different resolution settings is under a couple of percents. For the bright scenario the completeness is greater than $95\%$. For the faint survey case, the completeness is greater than 80\%.
This demonstrates how the sky lines impact the redshift completeness of an \OII spectroscopic survey. It shows it is necessary to have the smallest fiber possible to diminish the impact of the sky. The increase in resolution is not useful to cope with this problem. 

Finally the completeness in redshift is not driven by the resolution at the first order, but by the robustness of sky subtraction and the strength of the \OII flux. To obtain a precise estimate of the impact of sky lines on the redshift distribution completeness, a full end-to-end simulation is needed.

\subsection{Integrated velocity profile}
In this study, the integrated velocity profile of each galaxy within a fiber is assumed to be Gaussian, although galaxy rotation may create complications. Current data is not sufficient to explore this particular difficulty. Nearby galaxies are not representative of the properties of these higher-redshift galaxies, and surveys like MASSIV are limited to a sample of only 50 galaxies in the redshift range $0.6 < z < 1.6$ \citet{2012A&A...539A..92E}.

\subsection{Emission line flux ratio}
\label{subsec:fluxratio}
The flux ratio between the forbidden fine structure \OII lines varies with the surrounding electronic density between 0.35 (high electron density limit) and 1.5 (low density limit) \citep{2006MNRAS.366L...6P}. 
A precise estimation of the distribution of this ratio at $z\sim1$ has not been measured, although observations show the ratio does not take the extreme values 0.35 or 1.5, but seems to stay around 1.
A ratio of one is the best for separating the doublet. A different ratio can only decrease the efficiency at recognizing the doublet. Also this effect is symmetric, a ratio of 0.7 or 1.4 implies the loss of the same amount of doublets. We quantify this effect by varying the flux ratio of the lines simulated between 0.7 and 1.
For emission lines with total flux of $10^{-16} \mathrm{erg \, cm^{-2} \, s^{-1}}$ (DESi-like), a flux ratio of 0.7 (or 1.4) induces a decrease in the amount of doublets seen of 8.3\% at $R\sim4\,500$. The total number of doublets detected at $3\sigma$ goes from $\sim25\%$ to $\sim22.9\%$.
For emission lines with total flux of $3\times10^{-16} \mathrm{erg \, cm^{-2} \, s^{-1}}$ (SDSS-IV/eBOSS-like), a flux ratio of 0.7 (or 1.4) induces a decrease in the amount of doublets seen of 9.1\% at $R\sim3\,300$. The total number of doublets detected at $3\sigma$ diminishies from $\sim90\%$ to $\sim81.8\%$.

\section{Conclusion}
\label{sec:ccl}
Large spectroscopic redshift surveys are being designed to measure galaxy redshifts using the \OII emission line doublet and trace the large-scale matter distribution. This study shows we should be optimistic regarding their feasibility.
We have shown how the observation of the doublet evolves with the instrumental resolution and the line velocity dispersion. Also, we quantified the impact of sky lines on the redshift completeness of such a survey.

In light of the numbers obtained, we foresee two strategies about the choice of the resolution for future spectrographs:
For bright \OII emitter surveys (like SDSS-IV/eBOSS), a resolution of $R\sim2\,500$ (current SDSS spectrograph) is sufficient to obtain a fair sample of doublets (60\%) in order to train the pipeline to recover all the \OII redshifts. Increasing the resolution to 3$\,$300 allows to get 90\% of doublets. For a small increase in resolution, the redshift determination efficiency doubles. The impact of the sky lines on the completeness in redshift is smaller than $6\%$.
For faint \OII emitter surveys (like DESi), we recommend to push the resolution to the highest. Knowing there is a limited number of pixels on the detector (4k), and that the highest resolution possible on a three channel spectrograph is $R\sim4\,500$ at $7\,500\AA$, to go beyond, it is necessary to use a four channel spectrograph. Practically with a resolution of 4$\,$500, one would obtain 25\% of doublets, which is enough to train the pipeline to assign correct redshift.


\begin{acknowledgements}
JPK acknowledges support from the ERC advanced grant "LIDA". 
This work was supported by the United States Department of Energy Early Career program via grant DE-SC0003960 and by the National Science Foundation via grant AST-0806732.
\end{acknowledgements}

\bibliographystyle{aa}
\bibliography{biblio.bib}

\end{document}